\documentclass[%
 reprint,
 showpacs,
 amsmath,amssymb,
 aps,
]{revtex4-1}

\usepackage{epsfig,graphicx}% Include figure files
\usepackage{dcolumn}% Align table columns on decimal point
\usepackage{bm}% bold math
\usepackage{dcolumn}% Align table columns on decimal point

\begin{document}

\title{Results of a search for 2$\beta$-decay of $^{136}$Xe with high pressure copper proportional counters in Baksan Neutrino Observatory INR RAS}

%--------------------------------------------------

\author{
Ju.M.~Gavriljuk$^a$,
A.M.~Gangapshev$^{a}$, %\\
V.V.~Kazalov$^{a}$,
V.V.~Kuzminov$^{a}$, \\
S.I.~Panasenko$^{b}$,
S.S.~Ratkevich$^{b}$,
D.A.~Zhantudueva$^{a}$,
S.P.~Yakimenko$^{a}$
\\
$^a$ \small{\em Baksan Neutrino Observatory INR RAS} \\
\small{\em pos.~Neitrino, Elbrus raion, 361609 Kabardino-Balkaria, Russia} \\
$^b$ \small{\em Karazin Kharkiv National University, Ukraine} \\
\small{\em Svobody Sq. 4, 61022, Kharkiv. Ukraine}
}
\date{May 26, 2011}

\begin{abstract}
Search for $\beta\beta$-decay of $^{136}$Xe with two high pressure proportional counters is carried out in Baksan Neutrino Observatory. The experiment is based on comparison of spectra measured with natural and enriched xenon. The measured half life is equal to
T$_{1/2}=5.5^{+4.6}_{-1.7} \cdot 10^{21}$ yr (67\% C.L.)
for $\beta\beta2\nu$ decay mode. No evidence has been found for neutrinoless $\beta\beta$-decay. The decay half life limit based on data measured during 17280 h is T$_{1/2} (\beta\beta0\nu) \geq 4.9 \cdot 10^{23}$ yr (90\% C.L.).
\end{abstract}

%%%%%%%%%%%%%%%%%%%%%%%%%%%%%%%%%%%%%%%%%%%%%%%%%%%%%%%%%%%%%%%%%%%%%%%%%%%%%%%%%%


\maketitle

\section{Introduction}

The experimental investigation of the $\beta\beta$-decay of $^{136}$Xe has been started more then 20 years ago. But both two neutrino and neutrinoless modes of this process for $^{136}$Xe was not observed. The results of last experiments are presented in Table \ref{tab1}. The theoretical estimations of half lifetime for $\beta\beta2\nu$-decay are presented in Table \ref{tab2}. It is necessary to mention that in \cite{4} only one spectrum (measured with enriched $^{136}$Xe) was obtained. To calculate their limit it was assumed that at any effect/background ratio in the energy range under investigation the effect did not exceed the actually measured background increased by a systematic error given in $\sigma$ units ($\sigma$ is a standart deviation). In our work the measurements were performed with both enriched xenon and natural xenon simultaneously.
\begin{table*}
\caption{The results of some experiments for the search of $\beta\beta$-decay of $^{136}$Xe}. \label{tab1} \centering
\begin{tabular}{c|c|c}
%{\textwidth}{c|c|c}
\hline
Experiment & T$_{1/2}$ ($\beta\beta2\nu$), yr & T$_{1/2}$ ($\beta\beta0\nu$), yr\\
\hline
\hline
Gran Sasso \cite{1} & $\geq$1.6$\cdot$10$^{20}$
(95$\%$ C.L.) & $\geq$1.2$\cdot$10$^{22}$ (95\% C.L.)
\\
%\hline
GOTTHARD \cite{2} & $\geq$3.6$\cdot$10$^{20}$
(90$\%$ C.L.) & $\geq$4.4$\cdot$10$^{23}$ (90\% C.L.)
\\
%\hline
BNO INR RAS \cite{3} & $\geq 8.5 \cdot 10 ^{21}$
(90\% C.L.) & $\geq 3.1 \cdot 10 ^{23}$ (90\% C.L.)
\\
%\hline
DAMA/LXe \cite{4} & $\geq$1.0$\cdot$10$^{22}$
(90\% C.L.) & $\geq$1.2$\cdot$10$^{24}$ (90\% C.L.)
\\
\hline
\end{tabular}
\end{table*}

%\vspace{1cm}

\begin{table*}
\caption{The theoretical estimations for $\beta\beta2\nu$-decay of
$^{136}$Xe}. \label{tab2} \centering
\begin{tabular}{c|c}
\hline
Authors & T$_{1/2}$  ($\beta\beta2\nu$), yr\\
\hline
\hline
E. Caurier et al. \cite{5} & 2.1$\cdot$10$^{21}$ \\
%\hline
O.A. Rumyantsev, M.G. Urin \cite{6} & 1.0$\cdot$10$^{21}$
\\
%\hline
A. Staudt et al. \cite{7} &
1.5$\cdot10^{19}\div2.1\cdot10^{22}$
\\
%\hline
P. Vogel and M.R. Zirnbauer \cite{8} &
1.5$\cdot10^{20}\div1.5\cdot10^{21}$
\\
\hline
\end{tabular}
\end{table*}

\section{Experimental setup}

\begin{figure}%[pt]
%\begin{center}
\includegraphics*[width=8cm]{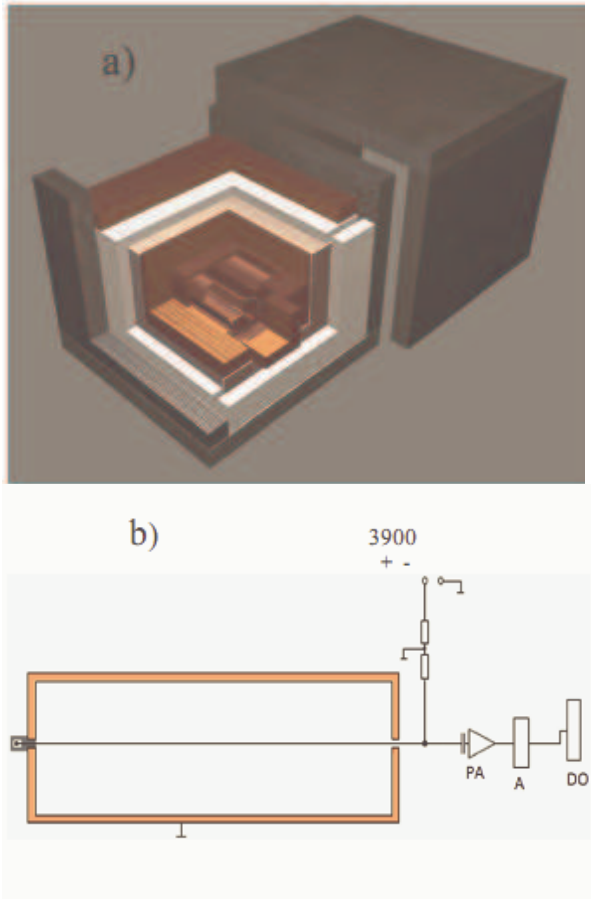}
\caption{\label{inst} a) - the schematic view of the installation and  b) - the electric scheme of installation,
PA - preamplifier, A - amplifier, DO - digital oscilloscope.}
\end{figure}

The measurement was carried out with two high pressure copper proportional counters (CPC $N$1 and CPC $N$2). While one of them is filled with enriched xenon (93$\%$ of $^{136}$Xe), the other one is filled with natural xenon depleted by light isotopes (9.2$\%$ of $^{136}$Xe). Both CPC-s are surrounded by passive shield consisting of 20 cm of copper (keeped in underground facilities over 20 yr), 8 cm of borated polyethylene and 23 cm of lead (see Fig.\ref{inst}a).
\begin{figure}%[pt]
\includegraphics*[width=7cm,,angle=0.]{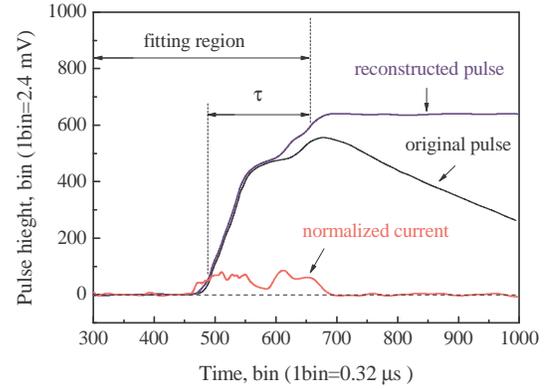}
\caption{\label{pulse} The samples of the pulses (initial pulse,
reconstructed pulse and normalised pulse).}
%\end{center}
\end{figure}
The installation is located in the deep underground laboratory of the BNO INR RAS at the depth 4900 m w.e,
where the flux of muons is decreased by factor 10$^7$ and evaluated as 2.23$\cdot 10 ^{-9} cm ^{-2} s ^{-1}$ \cite{9}.
The parameters of the CPC-s are: working pressure - 14.8 bar, fiducial volume - 8.39 l, biased voltage - 3900 V.
Signals were read out from the end of anode wire through charge sensitive preamplifier (PA). Then they were supplied to the digital
oscilloscope through amplifier (see Fig.\ref{inst}b).

To exclude the influence of PA capacity charge decay the pulses were reconstructed by software (taking into account this decay). The amplitudes of the reconstructed pulses (A) were used to construct the energy spectrum. For detailed analysis the following pulse shape parameters was used: the pulse rise time ($\tau$) and parameter $\delta$ defined as:
\begin{equation*}
    \delta =1000 \cdot(1-\frac{\sum{(y_{i}-g_{i})^{2}}} {\sum{(y_{i}-\overline{y})^{2}}}),
\end{equation*}
where the $y_{i}$ is value of normalized current at the point \emph{i}, $g_{i}$ - value of gaussian at the same point,
$\overline{y}$ - mean value of the normalized current in fitting region. The region of fitting by gaussian is
restricted by point of pulse beginning and point where the reconstructed pulse get a value 0.9$\cdot$A. In Fig.\ref{pulse}
samples of the pulses are shown.

\section{Data treatment}

From previous experiments \cite{3} it was seen that significant part of the counters background is due to $\alpha$-particles produced in gas volume. To define parameters of $\alpha$-particle events the measurement of background of one of CPC-s filled with xenon with admixtures of $^{222}$Rn ($\alpha$-particles with energy of 5.49 MeV, 6.02 Mev and 7.69 Mev) was performed. The results of this measurement are presented in \cite{10}.
\begin{figure}[pt]
\includegraphics*[width=6cm]{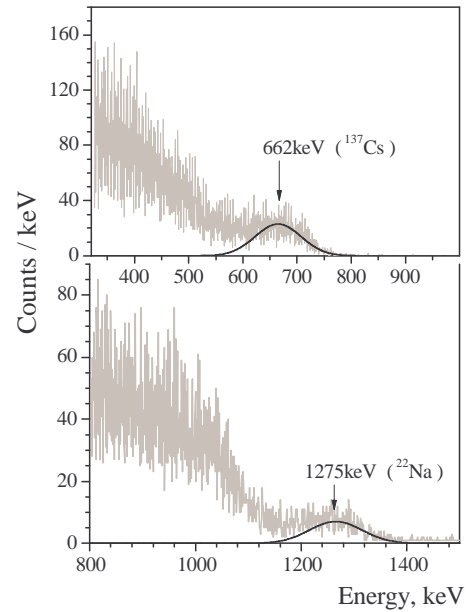}
\caption{\label{calibr} The calibration spectra of $^{137}$Cs and $^{22}$Na sources.}
\end{figure}

Main measurements performed in 4 runs. In the first run CPC $N1$ was filled with natural xenon and CPC $N2$ with enriched xenon. Duration of the run was $\geq$4000 h. Then first CPC was refilled with enriched xenon, second one with natural xenon, after each run the refilling procedure was repeated. Such a procedure allows eliminate systematic error from possible differences between counters.To exclude the contribution of $^{222}$Rn (which goes from gas purification system) to the CPC-s background the data of the first 500 h of measurements were not used for analysis. The calibration of the counters was carried out every $\sim$4 weeks ($\sim$700 h). The samples of calibration are presented in Fig.\ref{calibr}.

\begin{figure}[pt]
\begin{center}
\includegraphics*[width=7cm]{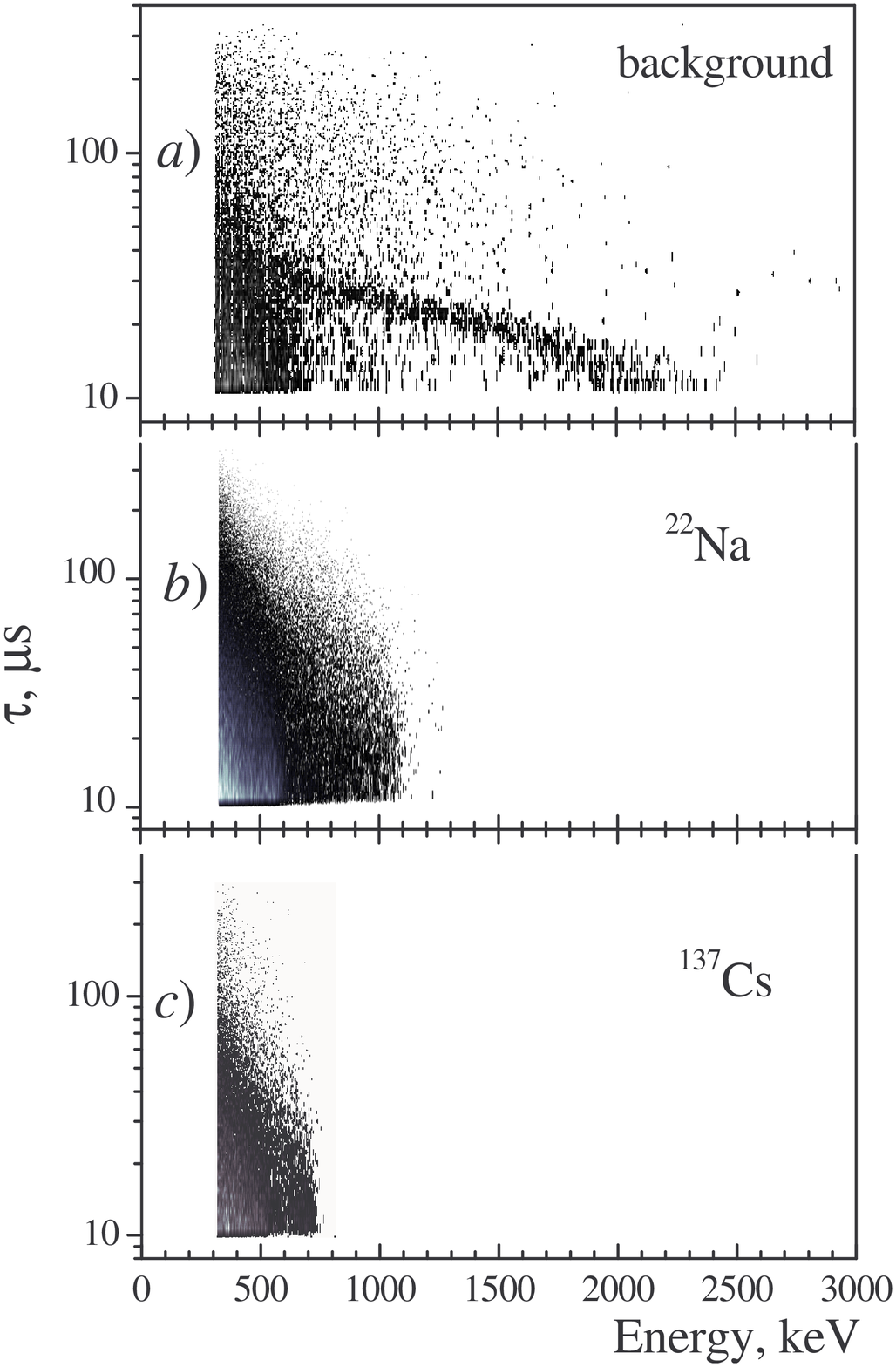}
\caption{\label{et} Distribution of events versus energy and
$\tau$ - \emph{a},\emph{b} and \emph{c} for the background ($\Delta t=3300$ h), $^{22}$Na- and
$^{137}$Cs-source respectively.}
\end{center}
\end{figure}

\begin{figure}[pt]
\begin{center}
\includegraphics*[width=7cm]{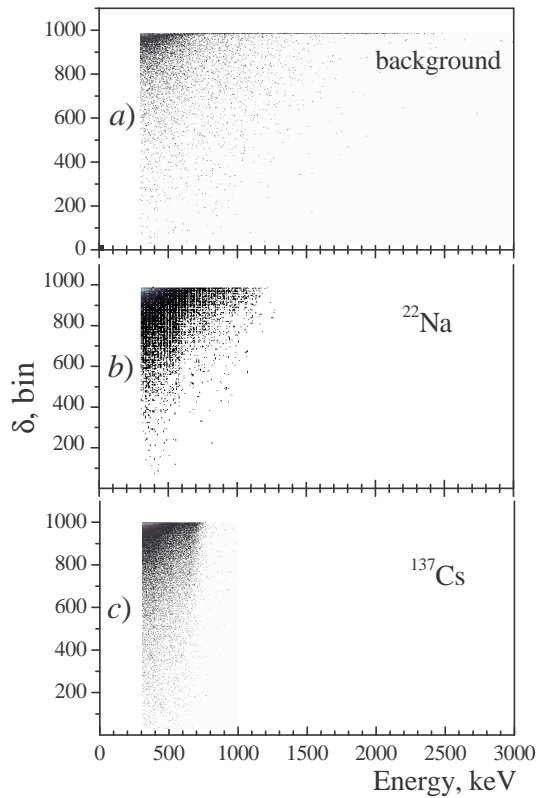}
\caption{\label{ed} Distribution of events versus energy and
$\delta$ - \emph{a},\emph{b} and \emph{c} for the background ($\Delta t=3300$ h), $^{22}$Na- and
$^{137}$Cs-source respectively.}
\end{center}
\end{figure}

\begin{figure}[pt]
\includegraphics*[width=7cm]{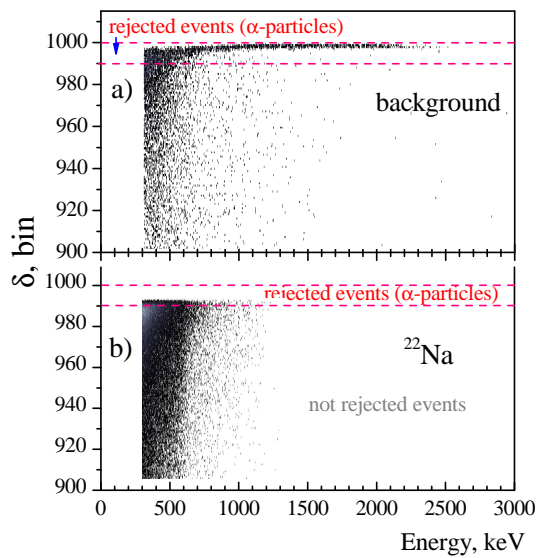}
\caption{\label{ev_cut}Distribution of the background events and $^{22}$Na-source versus energy and $\delta$ in the region for $\delta>900$  (cut of the events with $\delta>990$ is shown).}
\end{figure}

Distribution of the background events (3300 h of measurements), $^{22}$Na and $^{137}$Cs source events versus energy-$\tau$ and energy-$\delta$ are presented in Fig.\ref{et}(\emph{a}-\emph{c}) and Fig.\ref{ed}(\emph{a}-\emph{c}) respectively.
It is seen that distribution of the background events differ from $^{22}$Na events significantly. This difference is clearly seen in distribution of events versus energy and $\delta$ for energy greater then 800 keV (see Fig.\ref{ev_cut}.).
\begin{figure}%[hp]
\includegraphics*[width=7.0cm]{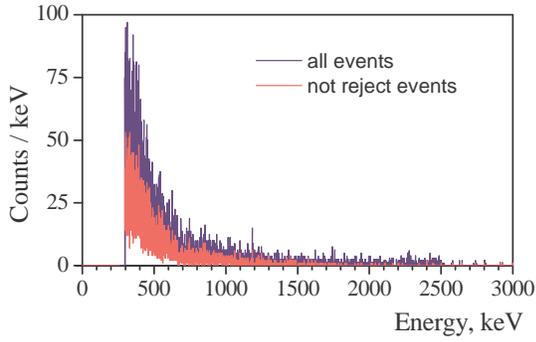}
\caption{\label{n_e_sel} The energy spectra measured during 4900 h, blue curve - before rejection of pulses with $\delta > 990$, orange curve - after rejection.}
\end{figure}
It was used to separate $\alpha$-particle events and electron events (events from $\alpha$-particles have $\delta >990$, most events from electron have $\delta\leq990$). The transformation of the background spectrum after such rejection is seen in Fig.\ref{n_e_sel}.
%\newpage

\begin{figure}[pt]
\includegraphics*[width=8.0cm]{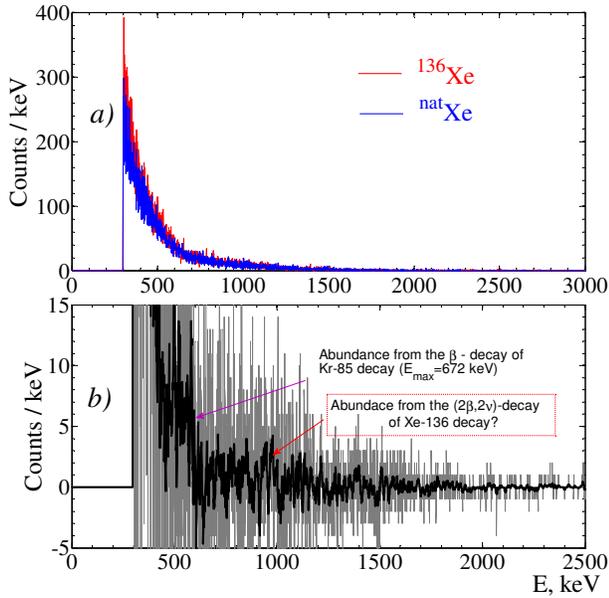}
\caption{\label{n_e_tot} \emph{a}) - the energy spectra measured during 13200 h and \emph{b}) - difference in spectra between enriched and natural xenon.}
\end{figure}

\section{Results}

The effect from $\beta\beta2\nu$-decay of $^{136}$Xe is determined with comparison of total spectra measured with enriched and natural xenon (see Fig.\ref{n_e_tot}).
Number of events in the energy region $0.7\div2.48$ MeV registered by CPC \emph{N}1 and CPC \emph{N}2 at 3300 h in each run of measurements are presented in Table \ref{tab3}.
\begin{table}
\caption{The number of events with energy 0.7$\div$2.48 MeV,
registered in CPC $N$1 and CPC $N$2 during 3300 h of measurements
for each run.} \label{tab3} \centering
\begin{tabular}{c|c|c}
\hline
run & CPC \emph{N}1 & CPC \emph{N}2 \\
\hline
\hline
1 & 2396 ($^{nat}$Xe) & 2591 ($^{136}$Xe) \\
%\hline
2 & 1953 ($^{136}$Xe) & 1968 ($^{nat}$Xe) \\
%\hline
3 & 1259 ($^{136}$Xe) & 1224 ($^{nat}$Xe) \\
%\hline
4 & 1383 ($^{nat}$Xe) & 1477 ($^{136}$Xe) \\
\hline
\end{tabular}
\end{table}
The evaluated effect is 309$\pm119(stat)\pm72(syst)$. The total deviation is $\sigma_{tot}=\sqrt{\sigma^{2}_{stat}+\sigma^{2}_{syst}}=139$.
So effect is $+2.22\sigma_{tot}$. Taking into account efficiencies and the different content of $^{136}$Xe in two samples we obtain the half lifetime for $\beta\beta2\nu$-decay:
\begin{eqnarray}
\nonumber
T_{1/2}(\beta\beta2\nu) &=& \frac{ln(2) \cdot t \cdot N \cdot \epsilon}{309\pm139}= \\
\nonumber
& & \\
\nonumber
                     &=& 5.5 ^{+4.6} _{-1.7}\cdot10^{21} {\texttt{ yr (67\% C.L.)}}, \\
\nonumber
& &
\end{eqnarray}
where: $t=13200$ h $=1.507$ yr - measurements time, $N=3.06\cdot10^{24}$ - difference in quantity of $^{136}$Xe atoms in xenon samples, $\epsilon=0.535$ - efficiency after rejection of events with $\delta>990$ and $E<700$ keV.

\begin{table}
\caption{The number of events with energy 2312$\div$2646 keV,
registered in CPC \emph{N}1 and CPC \emph{N}2
for four series of measurements of the counter background.}
\label{tab4}
%\centering
\begin{tabular}{c|c|c|c}
\hline
Run &  Meas. time, hours  & CPC \emph{N}1 & CPC  \emph{N}2 \\
\hline
\hline
1 & 3300 & 2 ($^{nat}$Xe) & 0 ($^{136}$Xe) \\
%\hline
2 & 4170 & 6 ($^{136}$Xe) & 1 ($^{nat}$Xe) \\
%\hline
3 & 6410 & 1 ($^{136}$Xe) & 4 ($^{nat}$Xe) \\
%\hline
4 & 3400 & 3 ($^{nat}$Xe) & 1 ($^{136}$Xe) \\
\hline
\end{tabular}
\end{table}

To evaluate the $\beta\beta0\nu$-effect the energy spectra in region
2312$\div$2646 keV were analyzed.
The data for the analysis $0\nu$-mode have been taken for 17280 hours of measurements.
This energy region is determined from calculated energy resolution for 2479 keV electrons (R=7.0$\%$, $2\sigma=147$ keV) and systematic error in definition of peak position ($\pm 20$ keV). Number of events in the energy region $2312\div2646$ keV registered by CPC \emph{N}1 and CPC \emph{N}2 in each run of measurements are presented in Table \ref{tab4}.
Using recommendation given in \cite{11} and assuming that mean background is 10 events
and measured one is 8 events, we obtain:
\begin{equation}
\nonumber
    T_{1/2}(\beta\beta0\nu\geq\frac{ln(2) \cdot t \cdot N \cdot \epsilon}{4.22}=4.9\cdot10^{23} {\texttt{ yr (90\% C.L.),}}
\end{equation}
where $t=17280$ h $=1.97$ yr and $\epsilon=0.5$.

This work was supported by the RFBR (Grant No. 07-02-00344) and the Neutrino Physics Program of the Presidium of the Russian Academy of Sciences.

\end{document}